\documentclass{elsart}

\usepackage{harvard}

\usepackage{graphicx}

\usepackage{amssymb}

\def\sun{\hbox{$\odot$}}

\begin{document}

\begin{frontmatter}
\title{Grand Unification of AGN and the Accretion and Spin Paradigms}


\author[Meier]{David L. Meier\thanksref{dlm}} 

\thanks[dlm]{E-mail: dlm@cena.jpl.nasa.gov}

\address[Meier]{238-332, Jet Propulsion Laboratory, California Institute of Technology, 
4800 Oak Grove Drive, Pasadena, CA 91109}

\begin{abstract}

While attempts to unify certain classes of AGN using orientation and environmental
effects have been successful, it is widely recognized that intrinsic properties
of the accreting black hole system also must play a role in determining the
appearance of such an object.  In addition to mass and accretion rate, the
angular momentum (or spin) of the black hole can play a crucial role in 
determining the power of a relativistic jet that is generated by magnetohydrodynamic
acceleration near the hole.  In this paper a scenario is presented, based on
accretion theory and recent models of MHD jet production, in which the primary
(although not only) parameter differentiating between radio loud and quiet
objects is the black hole spin, and that determining quasar vs. radio galaxy
is the accretion rate.  A surprising number of desirable features result from
these simple concepts and the accompanying equations.  In addition, there are
several testable predictions that can determine whether this grand unification
scheme has further merit.

\end{abstract}

\end{frontmatter}


\section{Introduction:  Heredity {\it vs.} Environment}
\label{intro}
Recent attempts to unify certain classes of active galactic nuclei (AGN) 
using orientation and environmental effects have been rather successful 
\cite{ob82} \cite{am85} \cite{barthel89} \cite{up95} \cite{wj99}.  
However, it is widely recognized that intrinsic properties of 
the accreting black hole systems that are thought to power AGN must also 
play a role in determining the appearance of these objects.  While there 
is still much to be learned, we now know enough about accretion onto black 
holes and about the production of jets to begin to develop {\em grand} 
schemes that attempt to unify most classes of these objects.  

In accretion and jet production theory the principal parameters determining 
the appearance and behavior of the system are the black hole mass $M_H$, 
the mass accretion rate $\dot{M}$, and the black hole angular momentum $J$, 
expressed in dimensionless form as $m_9 \equiv M_H / 10^9 \, M_{\sun}$ 
(where $M_{\sun}$ represents one solar mass), $\dot{m} \equiv \dot{M} / 
\dot{M}_{Edd}$ (where $\dot{M}_{Edd} = 4 \pi G M_H /\epsilon \kappa_{es} c = 
22 \, M_{\sun} yr^{-1} \, m_9$ is the accretion rate that produces one Eddington 
luminosity for an efficiency $\epsilon = 0.1$ and electron scattering opacity 
$\kappa_{es}$), 
and $j \equiv J/J_{max}$ (where $J_{max} = G M_H^2 / c$ is the angular momentum 
of a maximal Kerr black hole).  For most AGN and quasar models typical 
ranges of the parameters are $10^{-3} < m_9 < 10$, $10^{-5} < \dot{m} < 1$, 
and $0 < j < 1$.  While all parameters will affect the properties of an AGN 
to a certain extent, the purpose here is to identify the principal observable 
effects of each.

The unification through intrinsic effects discussed here is not meant to 
replace that which has been accomplished already through extrinsic 
considerations.  Rather, the accreting black hole system should be considered 
as providing the basic 
parent ensemble of AGN in which orientation and environmental effects 
can take place.  Pursuant to the anthropomorphological theme of this 
conference, this paper will deal with the ``psychology'' of AGN:  some 
traits are inherited from birth ({\it i.e.}, from the accreting system) 
and some are a product of the source's environment.  Considerations of 
both types of effects will lead to a better understanding of the life 
cycles of AGN.

\section{The Accretion Paradigm:  Effects of Black Hole Mass and Accretion 
Rate}
\label{accretion}
The accretion paradigm states that most, and perhaps all, AGN are powered 
by accretion onto a supermassive black hole \citeaffixed{br91}{see, {\it e.g.},}.  
Within this model 
$\dot{m}$ plays the most important role, determining the emission properties, 
and therefore the appearance, of the central source.  Objects with high 
accretion rate ($\dot{m} \gtrsim 0.1$) appear as an ``optical'' quasar
(of course, equally bright, if not brighter, in X-rays as well), while 
low sub-Eddington accretion ($\dot{m} \lesssim 10^{-2}$) produces a weak 
``radio'' core with substantially less optical emission.  A zero 
accretion rate produces a ``dead'' quasar --- a black hole detectable only 
through its gravitational influence on the galactic nucleus.  For a given 
$\dot{m}$ level, the black hole mass determines mainly the luminosity 
scaling.  

\subsection{Observational and Theoretical Evidence for the Accretion Paradigm}

There is both observational and theoretical evidence for this picture.  Based
on optical spectra of AGN, \citeasnoun{jw99} identify two classes of object:  
Class A (quasars, ``N'' radio galaxies, Seyfert galaxies) with strong {\em 
narrow} emission lines and Class B (most radio galaxies and weak radio cores) 
with weak or no narrow line emission.  Narrow line strength is considered 
a better parameter than, for example, broad line strength, as it is less 
affected by orientation effects \citeaffixed{am85}{see, {\it e.g.},}.
In this scheme Class A objects are 
identified with high nuclear gas content and, therefore, with high accretion 
rate, while Class B AGN are identified with low accretion rate.  An important 
point noted by Jackson \& Wall is that the Fanaroff \& Riley Class I radio 
sources are an homogeneous class, while FR II sources are not:  FR Is are 
all Class B AGN while most, but not all, FR IIs are Class A.  {\em That is, 
some FR II sources appear to have powerful jets and yet a rather low accretion rate.}

In addition, in present theories of accretion, a rapidly-accreting supermassive 
black hole embedded in an elliptical bulge of stars is predicted to appear as 
a quasar-like object (1-2 orders of magnitude brighter than its host).  
For $\dot{m} \gtrsim 0.1$ the standard 
disk models of \citeasnoun{ss73} appear to be most appropriate, with the optical 
luminosity (integrated disk emission from plasma at a temperature of $< 10^5 \, K$) 
scaling in solar units as
\begin{equation}
L^{opt}_{acc} \; \geq \; 1.7 \times 10^{12} L_{\sun} \, m_{9}^{1.27} \, 
\left( \frac{\dot{m}}{0.1} \right)^{0.6}
\end{equation}
If the black hole mass is directly related to galactic bulge luminosity, 
{\it i.e.}, $L^{opt}_{gal} \approx 2.4 \times 10^{10} L_{\sun} \, m_{9}^{0.8}$, 
as suggested by \citeasnoun{kr95}, then the ratio of accretion disk to bulge 
luminosity is
\begin{equation}
\frac{L^{opt}_{acc}}{L^{opt}_{gal}} \; > \; 68 m_{9}^{0.47}
\end{equation}
for $\dot{m} > 0.1$.  On the other hand, a drop in accretion rate well below this 
value produces a much fainter ``advection-dominated'' \cite{n98} accretion disk 
with bolometric luminosity dropping as $L^{bol}_{acc} \propto \dot{m}^2$, and 
optical luminosity dropping even faster than that.  (The exact accretion rate at which 
this drop-off occurs [$\dot{m}_{cr} \sim \alpha^2$] varies steeply with the value 
of the viscosity parameter $\alpha$, which is usually taken to be $\sim 0.3$
for ADAF models.)
Such disks are geometrically thick, optically 
thin, and emit mainly nonthermal radio emission.  For the remainder of this 
paper, we will take $\dot{m} = 0.1$ to be typical of Class A quasar-like objects 
and $\dot{m} = 0.01$ to be typical of Class B radio cores.

\subsection{Observations Explained}

Several interesting corollaries immediately follow from this picture.  
For black holes in elliptical galaxies, only dwarf ellipticals are brighter than 
their central quasar ($L^{opt}_{acc} < L^{opt}_{gal}$ when $m_{9} < 10^{-4}$, or 
when the galaxy magnitude is $\mathcal{M}^{opt}_{gal} > -13$).  However, for holes 
in spiral galaxies, where the optical luminosity is dominated by a stellar disk 
of luminosity $\sim 2 \times 10^{10} L_{\sun}$, the object will appear as a 
Seyfert galaxy rather than a quasar ($L^{opt}_{gal} > L^{opt}_{acc}$) for black 
holes of rather high mass ($m_{9} < 0.03$ or $M_{H} < 3 \times 10^{7} M_{\sun}$).
The simple accretion paradigm, therefore, accounts for several of the basic 
{\em optical} properties of Seyfert, radio galaxy, and quasar (parent) populations.

\section{The Spin Paradigm:  Effects of Black Hole Rotation}
\label{spin}

The spin paradigm states that, to first order, it is the normalized black 
hole angular momentum $j$ that determines whether or not a strong radio jet 
is produced \cite{wc95} \cite{bland99}.  If correct, this hypothesis has 
significant implications 
for how we should view the jets and lobes in radio sources:  the jet radio and 
kinetic energy comes directly from the rotational energy of a (perhaps 
formerly) spinning black hole.  {\em Radio sources are not powered (directly) 
by accretion.}

\subsection{Theoretical Arguments for the Spin Paradigm}

There is significant theoretical basis for this paradigm as well.  Several 
models of relativistic jet formation \cite{bz77} \cite{pc90} indicate 
that the jet power should increase as the square of the black hole 
angular momentum
\begin{equation}
L_{jet} \; = \; 10^{48} {\rm erg \, s^{-1}} \, \left( \frac{B_{p}}{10^5 G} \right)^2 
\, m_{9}^2 \, j^2
\end{equation}
where $B_{p}$ is the strength of the poloidal (vertical/radial) magnetic 
field threading the ergospheric and horizon region of the rotating hole.  
In this model rotational energy is extracted via a Penrose-like process:  
the frame-dragged accretion disk is coupled to plasma above and outside the 
ergosphere via the poloidal magnetic field;  some plasma is pinched and 
accelerated upward while some disk material is diverted into negative 
energy (retrograde) orbits inside the ergosphere, removing some of the 
hole's rotational energy.  The key parameter determining the efficiency 
of this process is the strength of the poloidal magnetic field.  The 
standard approach \citeaffixed{ms96}{{\it e.g.,}} to estimating $B_{p}$ 
is to set it equal to $B_{\phi}$, the dominant azimuthal magnetic field 
component given by the disk structure equations, yielding
\begin{eqnarray}
\label{eq_ljet_adaf}
L_{jet,B} & = & 2 \times 10^{45} {\rm erg \, s^{-1}} \, m_{9} \, \left( 
\frac{\dot{m}}{0.01} \right) \, j^2
\\
\label{eq_ljet_std}
L_{jet,A} & = & 3 \times 10^{49} {\rm erg \, s^{-1}} \, m_{9}^{1.1} \, \left( 
\frac{\dot{m}}{0.1} \right)^{0.8} \, j^2
\end{eqnarray}
for Class B (radio galaxy/ADAF) and Class A (quasar/standard disk) objects, 
respectively.  Note that, while the jet is not accretion-powered in this 
model, the efficiency of extraction is still essentially linear in $\dot{m}$.

\citeasnoun{lop99} have pointed out that taking $B_{p} \approx B_{\phi}$ may greatly 
overestimate the jet power from this process.  Using dynamo arguments they 
propose that a more realistic estimate for the equilibrium poloidal 
magnetic field is 
\begin{equation}
\label{eq_bpol}
B_{p} \approx (H/R) \, B_{\phi}
\end{equation}
where $(H/R)$ is the 
ratio of disk half-thickness to radius in the jet acceleration region.  For 
{\em thin} disks this yields a jet power of only 
$L_{jet} = 4 \times 10^{44} {\rm erg \, s^{-1}} \, m_{9}^{1.1} \, (\dot{m}/0.1)^{1.2} \, j^2$ 
--- less than the observed {\em radio} power of the strongest sources and 
much less than their inferred total jet power 
\citeaffixed{bicknell95}{see Bicknell, these proceedings, and}.  
However, there are several reasons for believing that even with equation 
(\ref{eq_bpol}) the field still can be quite large in many cases, and the 
jet power still comparable to equations (\ref{eq_ljet_adaf}) and (\ref{eq_ljet_std}).  
Firstly, for advective disks (both the accretion-starved kind [$\dot{m} << 1$] 
and the super-Eddington kind [$\dot{m} \gtrsim 1$]) the disk is geometrically 
thick ($H/R \sim 1$), yielding $B_{p} \approx B_{\phi}$ even within the dynamo 
argument.  Thick disks also can occur for an even broader range of accretion 
rate when the hole and disk spin axes are misaligned:  because of the Lens-Thirring
effect, 
the gas follows inclined orbits that do not close, creating shocks and dissipation that 
bloats the disk into a quasi-spherical, inhomogeneous inflow \cite{bland94}.  Furthermore, 
even when $H << R$, inside the last stable orbit (or in any other region 
of the disk where the infall velocity suddenly approaches the free-fall speed) 
conservation of mass will cause a drop in density and pressure.  The toroidal 
field may then be dynamically important, buckling upward out of the plunging 
accretion flow, resulting in $B_{p}$ being comparable to $B_{\phi}$ 
\cite{krolik99}.  

\begin{figure}
\begin{center}
\includegraphics*[width=11cm,angle=-90]{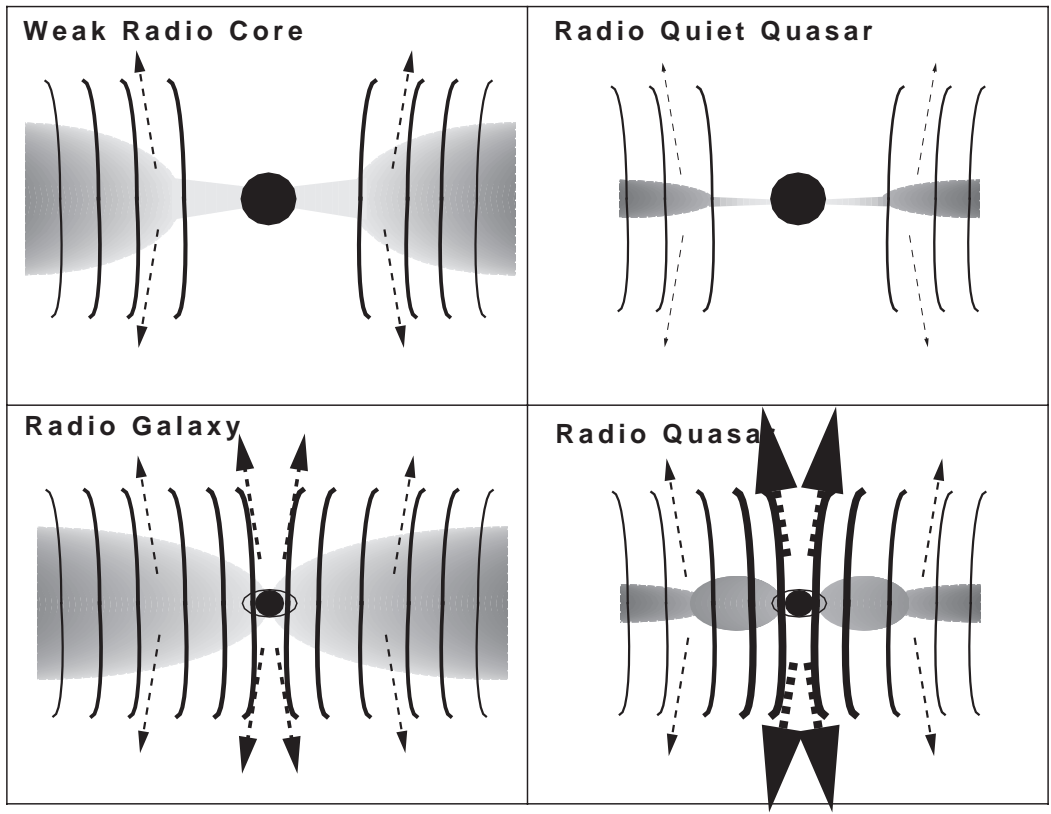}
\end{center}
\caption{Schematic representation of four possible combinations of $\dot{m}$ 
and $j$, drawn roughly to scale.  
The horizon interior to the hole is black, while the boundary of the 
ergosphere (the ``static limit'') is 
represented by an ellipse $0.5 \times 1.0$ Schwarzschild radius in size.  
Top panels depict non-rotating, Schwarzschild holes ($j \rightarrow 0$), 
bottom panels Kerr holes ($j \rightarrow 1$).  
Left panels show low accretion rate (ADAF) tori, right panels high accretion 
rate standard disk models.  
However, in the lower right panel the region of the disk 
experiencing significant frame dragging is bloated ({\it c.f.} Blandford 1994).
Widths of poloidal magnetic field lines and jet arrows are proportional to the 
logarithm of their strength.}  
\label{fig_schematics}
\end{figure}

Figure \ref{fig_schematics} summarizes the main features of the accretion and 
spin paradigms and shows the four possible combinations of high and low 
accretion rate and black hole spin.  It is proposed that these states 
correspond to different radio loud and quiet quasars and galaxies.  In the 
figure poloidal magnetic field strengths are estimated from equation 
(\ref{eq_bpol}), but $(H/R)$ is of order unity for the low $\dot{m}$ cases, 
and also for the high $\dot{m}$ Kerr case due to Lens-Thirring bloating of 
the inner disk.  Otherwise $(H/R)$ is calculated from the electron 
scattering/gas pressure disk model of \citeasnoun{ss73}, and disk field 
strengths are computed from that paper or from \citeasnoun{n98}, as 
appropriate.  The logarithms of the resulting poloidal field strengths, and 
corresponding jet powers, are represented as field line and jet arrow widths.  
In the Kerr cases, the inner disk magnetic field is significantly enhanced 
over the Schwarzschild cases, due in part to the smaller last stable orbit 
(flux conservation) and in part to the large $(H/R)$ of the bloated disks.  
The high accretion rate, Schwarzschild case has the smallest field --- 
and the weakest jet --- because the disk is thin, the last stable orbit is 
relatively large, and the Keplerian rotation rate of the field there is much 
smaller than it would be in a Kerr hole ergosphere.
Enhancement of the poloidal field due to the buoyancy process suggested by 
\citeasnoun{krolik99} is ignored here because we find it not to be a factor 
in the simulations discussed below.  If it were important, 
then the grand scheme proposed here would have to be re-evaluated, as the 
effect could produce strong jets (up to the accretion luminosity in power) 
even in the plunging region of Schwarzschild holes.   Then even the latter 
would be expected to be radio loud as well 
($L_{jet} \sim 10^{43-46} {\rm erg \, s^{-1}}$).

\subsection{General Relativistic Simulations of Magnetized Disks in Kerr 
Geometry}

Recent work by S. Koide, the author, and K. Shibata and T. Kudoh on general 
relativistic 
simulations of the accretion of magnetized material by black holes is 
beginning to provide a clearer picture of the development and evolution of 
a black hole magnetosphere and the resulting jet.  Initial indications are 
that rapid rotation of the black hole {\em and} rapid infall of the 
magnetized plasma into this rotating spacetime both contribute to powerful, 
collimated, relativistic jet outflows.  We find no evidence for buoyant 
poloidal field enhancement in the plunging region and, therefore, no reason 
to expect Schwarzschild holes to have a jet any more powerful than that 
estimated in Figure \ref{fig_schematics}.  

\begin{figure}
\begin{center}
\includegraphics*[width=7.5cm,angle=-90]{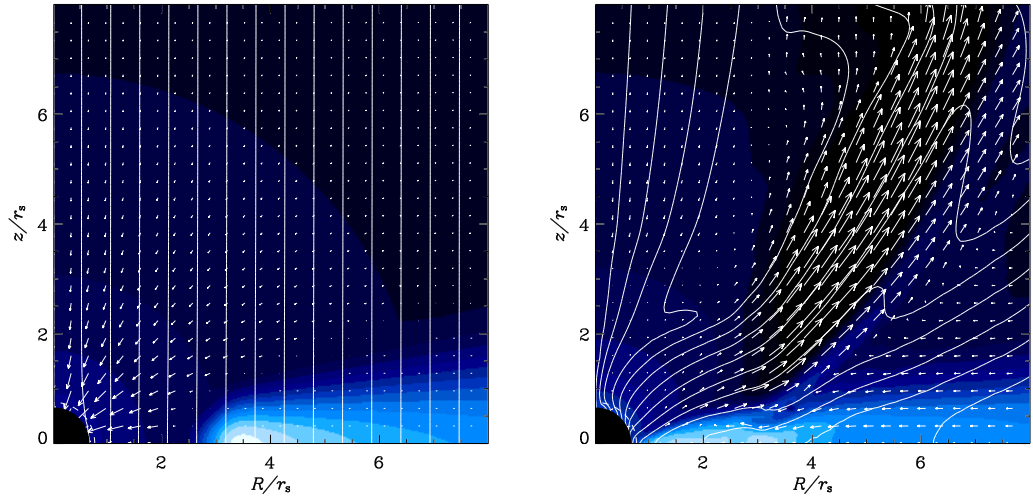}
\end{center}
\caption{General relativistic simulation of a magnetized flow 
accreting onto a Kerr black hole (after Koide {\it et al.} 1999a).
Left panel shows one quadrant of the initial model with the $j=0.95$ 
hole at lower left, freely-falling corona, and non-rotating disk. 
The hole rotation axis ($Z$) is along the left edge.  
The initial magnetic field (vertical lines) is weak compared 
to the matter rest energy density ($V_{A} = 0.01 c$).
Right panel shows final model at $t = 130 \, G M_{H} / c^3$. Some of 
the disk and corona have been accreted into the hole, threading 
the ergosphere and horizon with magnetic field lines that develop a 
significant radial component, and hence an azimuthal component as well, due 
to differential frame dragging.  The resulting 
jet outflow is accelerated by $\mathbf{J} \times \mathbf{B}$ forces 
to a Lorentz factor of $2.7$.}  
\label{fig_kgrmhd}
\end{figure}

Figure \ref{fig_kgrmhd} shows the initial and final state of a Kerr 
($j = 0.95$) black hole simulation with a weak 
($V_{A} \equiv B_{p} / \sqrt{4 \pi \rho_{disk}} = 0.01 c$) magnetic 
field and an inner edge at $R_{in} = 4.5 \, G M_H / c^2$.  For more 
details see \citeasnoun{koide99a}.  In this first simulation the disk 
was {\em non-rotating}, and so began to free-fall into the ergosphere 
(at $R_{ergo} = 2 GM_{H}/c^2$) as the calculation proceeded.  As with other 
MHD disk simulations \cite{ks95} \cite{ust95} \cite{op97} \cite{megpl97}, 
the field lines were wound up by differential rotation and a hollow jet 
of material was ejected along the poloidal field lines by the 
$\mathbf{J} \times \mathbf{B}$ forces; here the jet velocity was 
rather relativistic ($v_{jet} \sim 0.93c$  or $\Gamma \sim 2.7$).  However, 
as the disk was initially non-rotating, all the action was due to the 
differential dragging of frames by the rotating black hole.  Little or no 
jet energy was derived from the binding energy of the accreting material or 
its Keplerian rotation.

As a control experiment, the same simulation was also run with a Schwarzschild
($j=0$) hole.  The collapse developed a splash outflow due to tidal focusing 
and shocking of the inflowing disk, but no collimated MHD jet occurred.  

We also studied counter-rotating and co-rotating Keplerian disks with otherwise 
similar initial parameters \cite{koide99b}.  The counter-rotating case behaved nearly 
identically to the non-rotating case:  because the last stable retrograde 
orbit is at $R_{ms} = 9 G M_{H}/c^2$, the disk began to spiral rapidly 
into the ergosphere, ejecting a strong, black-hole-spin-driven MHD jet.
On the other hand, for prograde orbits $R_{ms} = G M_{H}/c^2$, so the 
co-rotating disk was stable, accreting on a slow secular time scale.
At the end of the calculation ($t_{max} = 94 \, G M_{H}/c^3$), when the 
simulation had to be stopped because of numerical problems, the disk had 
not yet accreted into the ergosphere and had produced only a weak MHD jet.
Future evolution of this case is still uncertain.  The Keplerian 
Schwarzschild case was previously reported by \citeasnoun{koide98} and 
developed a moderate sub-relativistic MHD jet similar to the aforementioned 
simulations of magnetized Keplerian flows around normal stars, plus a 
pressure-driven splash outflow internal to the MHD jet.  

In the Kerr cases we see significant magnetic field enhancement over the 
Schwarzschild cases due to 
compression and differential frame dragging.  It is this increase that 
is responsible for the powerful jets ejected from near the horizon. 
In the Schwarzschild cases, on the other hand,  --- particularly in the Keplerian disk 
case in \citeasnoun{koide98} ---  we do {\em not} see any MHD jet ejected 
from well inside $R_{ms}$.  That is, we do not see any increase in jet power that 
could be attributed to buoyant enhancement of the poloidal field in the 
plunging region.  While there is a jet ejected from inside the last stable 
orbit, it is pressure-driven by a focusing shock that develops in the accretion flow.
The magnetically-driven jet that does develop in the Schwarzschild case emanates 
from near the last stable orbit as expected, not well inside it.  It is therefore 
concluded that a powerful MHD jet will be produced near the horizon if and only 
if the hole is rotating rapidly.

\subsection{Observational Advantages of the Spin Paradigm}

There are several observational advantages to the spin paradigm for AGN.
It lifts the degeneracy of accretion onto black holes and explains the difference 
between radio loud and radio quiet objects:  a given central source type 
($m_{9}$, $\dot{m}$) can have a powerful jet ($j \rightarrow 1$) or 
little or no jet at all ($j \rightarrow 0$).  Maximal Kerr holes produce 
powerful radio sources while Schwarzschild holes produce ``radio 
quiet'' objects.  Even with only modest efficiency in converting the jet 
into particles and fields, the spent rotational energy of a black hole 
($E_{rot} \approx 10^{62} {\rm erg} \, m_{9} \, j^2$) can easily account 
for the observed energy in the lobes of the most powerful sources \cite{deyoung76}.
Taking the FR I/II break to occur at a jet power of $10^{44} {\rm erg} \, s^{-1}$ 
for a $10^9 M_{\sun}$ hole (using a Bicknell flux conversion 
factor of $\kappa_{\nu} = 10^{-11} {\rm Hz}^{-1}$) equations (\ref{eq_ljet_adaf}) 
and (\ref{eq_ljet_std}) 
predict correctly that radio quasars (Class A, $\dot{m}=0.1$, $j=0.01$-$1.0$) 
will be predominantly FR II sources ($10^{45-49} {\rm erg \, s^{-1}}$), 
low-spin radio galaxies (Class B, $\dot{m} = 0.01$, $j=0.01$-$0.1$) 
will be FR I sources ($10^{41-43} {\rm erg \, s^{-1}}$), but there 
should be a population of Class B sources with low accretion rate and high 
black hole spin ($\dot{m} = 0.01$, $j=0.1$-$1.0$) that produce some FR II 
sources without central quasars ($10^{43-45} {\rm erg \, s^{-1}}$).
These are the Class B FR II sources noted by \citeasnoun{jw99}.  

The spin paradigm even offers an explanation for why present-day giant 
radio sources occur only in elliptical galaxies and why the distribution 
of optically-selected quasars may be bi-modal in radio power.  The e-folding 
spindown time ($E_{rot}/L_{jet}$) is very short --- much shorter than a 
cosmic evolutionary time
\begin{equation}
\label{eq_tau_spindown}
\tau_{spindown} \; \approx \; 10^6 {\rm yr} \, \left( \frac{\dot{m}}{0.1} 
\right)^{-0.8} \, \left( \frac{\zeta_{duty}}{0.1} \right)^{-1}
\end{equation}
even with a jet active duty cycle of only 10\%.  As a 
result, {\em all AGN should be radio quiet at the present epoch}, their black 
holes having spun down when the universe was very young.  In order to 
continually produce radio sources up to the present epoch, there must be 
periodic input of significant amounts of angular momentum from accreting stars 
and gas, or from a merger with another supermassive black hole 
\citeaffixed{wc95}{see also}. Such activity is triggered most easily by 
violent events such as galaxy mergers.  Since only elliptical 
galaxies undergo significant mergers (the merging process is believed to be
{\em responsible} for their elliptical shape), {\em only ellipticals are 
expected to be giant radio sources in the present epoch}.  With little merger 
activity, spiral galaxies are expected to be relatively radio quiet.  Since 
merging and non-merging galaxies will fuel and re-kindle their black hole's 
spin in very different ways, bi-modality in the radio luminosity distribution 
is to be expected.

\section{An Example Grand Unification Scheme --- Observational Predictions}

\begin{figure}
\begin{center}
\includegraphics*[width=8cm,angle=90]{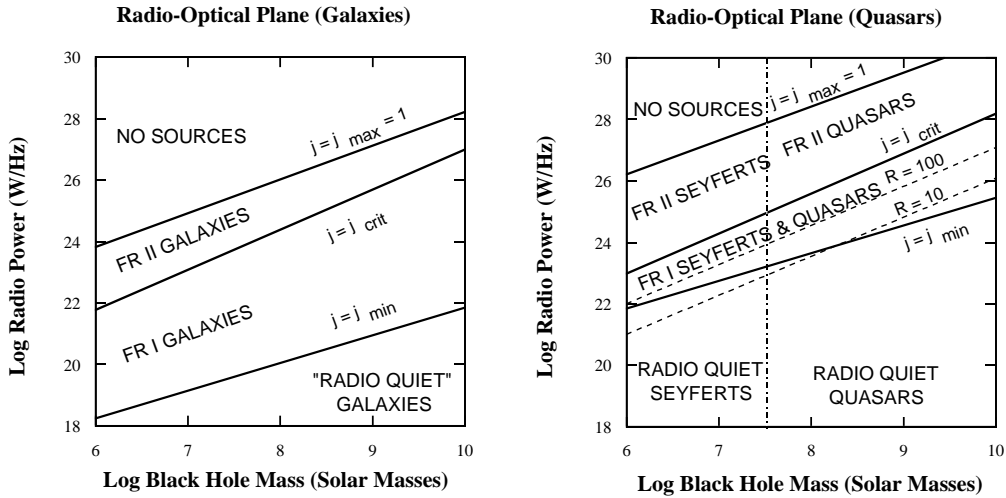}
\end{center}
\caption{Example of a grand unification scheme (after Meier 1999).
Left panel shows Class B 
objects ({\it e.g.}, radio galaxies; $\dot{m} = 0.01$), right 
panel Class A objects ({\it e.g.}, quasars; $\dot{m} = 0.1$).  Lines 
dividing the radio-optical planes are discussed in the text.}
\label{fig_grand_sch}
\end{figure}

Figure \ref{fig_grand_sch} shows two radio-optical planes for Class A and 
Class B objects in an example grand unification scheme 
\citeaffixed{meier99}{after}.  These figures are to be compared with observed 
radio-optical planes like that of \citeasnoun{lo96}.  As 
the optical luminosity of the galaxy or quasar scales with black hole mass in 
both cases, the horizontal axis is $M_{H}$ in both.  The vertical axis is 
the observed radio power using the Bicknell factor $\kappa_{\nu}$ to 
convert $L_{jet}$ to $P^{rad}$.  The curves separate the plane into several 
different black hole spin states:  NO sources ($j > 1$ is not allowed); 
FR II sources ($j_{crit} < j < 1$), FR I sources ($j_{min} < j < j_{crit}$), 
and radio quiet objects ($j < j_{min}$).  For $j_{crit}$ we have used here 
the generalized magnetic switch of \citeasnoun{meier99}, but Bicknell's FR I/II 
transonic condition could be substituted.  For $j_{min}$ we have chosen 
the point where the predicted black hole MHD power in equation (\ref{eq_ljet_std}) 
equals the thin disk MHD power with $B_{p} = (H/R) B_{\phi}$.  Other 
observational definitions of radio quietness (using the radio-optical flux 
ratio) are shown.

This grand unification scheme makes some interesting predictions.  There 
should be a population of sources corresponding to FR I quasars (high 
$\dot{m}$, low $j$).  These were formerly FR II sources, but their holes have 
since spun down. However, the spindown time for Class A sources is so short 
(equation \ref{eq_tau_spindown}) that it is likely that the FR II hot spots 
will still be radiating as the source goes through the FR I phase.  Such 
hybrid FR II sources could be identified by {\em young} diffuse emission or 
``bridges'' between the radio core and the hot spots.  The FR IIa quasars 
identified by R. Daly (these proceedings) are candidates for such sources;  
the FR IIa/IIb transition also appears to occur near the $j=j_{crit}$ line in 
Figure \ref{fig_grand_sch}b.

There also should be a population of high redshift, faint sub-mJy FR I and II 
radio sources associated with spiral galaxies or pre-spiral bulges ($M_{H} < 
10^7 M_{\sun}$).  If the accretion rate was high at that time --- the most 
likely case  in the early universe --- then these may appear as optically faint radio quasars 
($L^{opt} \lesssim 10^{43} {\rm erg \, s^{-1}}$, $P^{rad} \lesssim 10^{23-27} 
{\rm W \, Hz^{-1}}$).  Their numbers should be a significant fraction of the 
present-day spiral population, exceeding the contribution of quasars residing in 
elliptical galaxies alone.

\section{Discussion and Conclusions}

Considerably more theoretical and observational work is needed to further 
refine and test grand schemes of this type, including more detailed simulations of 
Kerr black hole magnetospheres and a better understanding of the 
structure and strength of the magnetic field near rotating (and non-rotating) 
holes.  Furthermore, any complete grand scheme must explain, from first 
principles, how the fundamental features of optical quasar spectra 
(``nonthermal'' continuum, broad and narrow lines) are produced and how 
accreting black holes should look at near-Eddington and super-Eddington 
accretion rates.

On the observational side, good methods of estimating the central accretion
rate and black hole spin are needed to locate each source in 
($M_{H}$, $\dot{M}$, $J$)-space.  As has been done in the past for stars 
and galaxies, when relations between fundamental parameters and observable 
properties are understood, an overall picture of how seemingly disparate 
objects are related begins to emerge.

\ack{
The author is grateful to S. Koide for discussions and for permission to use 
Figure \ref{fig_kgrmhd}.  
This research was carried out at the Jet Propulsion Laboratory, California
Institute of Technology, under contract to NASA.}



{\em Q:  P. Barthel:}  The PG (Palomar-Green) QSOs show bi-modality in radio loudness 
(Kellermann {\it et al.} 1989, AJ, 98, 1195).  If this holds up, will you be 
worried?  That is to say, would you expect a continuous distribution of radio 
loudness in optically-selected QSOs?

{\em A:  D. L. Meier:}  No, I wouldn't be worried; in fact, bi-modality is to be expected 
in a scenario where one population of objects is fueled by mergers (ellipticals) 
and one is fueled by mostly internal processes within the galaxy (spirals). 
One would expect these to have very different distributions in radio power. 
These issues also have been discussed by \citeasnoun{wc95}.

{\em Q:  D. E. Harris:}  Does your code accelerate particles (and if so, which), or 
is Poynting flux the primary output?

{\em A:  D. L. Meier:}  The general relativistic Kerr code of Dr. Koide is an ideal 
magnetohydrodynamic code.  There is only a single (fully ionized) fluid that is 
accelerated by the $\mathbf{J} \times \mathbf{B}$ force.  We do not compute 
a particle spectrum, unlike T. Jones and I. Tregillis, who will discuss their 
jet propagation simulations later in the meeting.


\end{document}